\documentclass[useAMS,usenatbib]{mn2e}
\usepackage{graphics,rotating,multirow,bm,color,hyperref}
\hypersetup{
    colorlinks=false,
    linkcolor=black,
    citecolor=black,
}
\usepackage{times,graphicx,amsmath,amsfonts,amssymb,epstopdf}
\usepackage[normalem]{ulem}
\usepackage[T1]{fontenc}

\title[Cluster gas fraction as a test of gravity]
  {Cluster gas fraction as a test of gravity}
\author[Li, He \& Gao]
  {Baojiu~Li$^{1,}$\thanks{E-mail: baojiu.li@durham.ac.uk},
  Jian-hua~He$^{2}$, Liang Gao$^{3,1}$\\
  $^1$Institute for Computational Cosmology, Department of Physics, University of Durham, South Road, Durham DH1 3LE, UK\\
  $^2$INAF-Observatorio Astronomico, di Brera, Via Emilio Bianchi, 46, I-23807, Merate (LC), Italy\\
  $^3$National Astronomical Observatories, Chinese Academy of Sciences, 20A Datun Road, Chaoyang District, Beijing 100012, China}

\pagerange{\pageref{firstpage}--\pageref{lastpage}} \pubyear{2015}

\def\LaTeX{L\kern-.36em\raise.3ex\hbox{a}\kern-.15em
    T\kern-.1667em\lower.7ex\hbox{E}\kern-.125emX}

\newcommand{\mpl}{M_{\rm Pl}}
\newcommand{\rd}{{\rm d}}

\begin{document}

\label{firstpage}

\maketitle

\begin{abstract}
We propose a new cosmological test of gravity, by using the observed mass fraction of X-ray emitting gas in massive galaxy clusters. The cluster gas fraction, believed to be a fair sample of the average baryon fraction in the Universe, is a well-understood observable, which has previously mainly been used to constrain background cosmology. In some modified gravity models, such as $f(R)$ gravity, gas temperature in a massive cluster is determined by the {\it effective} mass of that cluster, which can be larger than its {\it true} mass. On the other hand, X-ray luminosity is determined by the true gas density, which in both modified gravity and $\Lambda$CDM models depends mainly on $\Omega_{\rm b}/\Omega_{\rm m}$ and hence the true total cluster mass. As a result, the standard practice of combining gas temperatures and X-ray surface brightnesses of clusters to infer their gas fractions can, in modified gravity models, lead to a larger  -- in $f(R)$ gravity this can be $1/3$ larger -- value of $\Omega_{\rm b}/\Omega_{\rm m}$ than that inferred from other observations such as the CMB. A quick calculation shows that the Hu-Sawicki $n=1$ $f(R)$ model with $|\bar{f}_{R0}|=3\sim5\times10^{-5}$ is in tension with the gas fraction data of the 42 clusters analysed by \citet{allen2008}. We also discuss the implications for other modified gravity models.
\end{abstract}


\section{Introduction}

\label{sect:intro}

In recent years, attempts to understand the origin of the accelerated cosmic expansion \citep[][]{riess1998,perlmutter1999} have led to a large number of theoretical models \citep[][]{cst2006}. Apart from the current standard model, which assumes that the acceleration is caused by a cosmological constant $\Lambda$ (hence the name $\Lambda$-cold-dark-matter, or $\Lambda$CDM), these model can be roughly put in two categories: dark energy -- which replaces $\Lambda$ by some dynamical field, and modified gravity -- which assumes that there is no exotic matter species beyond the standard CDM model but gravity is not described by General Relativity (GR) on cosmological scales \citep[see, e.g.,][for some recent reviews]{clifton2012,joyce2015}. Some of the models in the latter class, such as the chameleon theory \citep{chameleon}, of which the well-known $f(R)$ gravity \citep{cddatt2005} is an example, have been active research topics in recent years.

Ultimately, any new cosmological model or theory of gravity should be put to rigorous tests against observational data. For this reason, a number of tests have been proposed or applied in the past to examine the viability of the models \citep[see, e.g.,][for some recent reviews in $f(R)$ gravity]{lombriser2014,demartino2015}. The present paper shall follow the same line to propose a test using observations of galaxy clusters.

Galaxy clusters are the largest gravitationally bound and virialised objects in our Universe. The most massive clusters observed today typically have masses in the range of $\sim10^{14}-10^{15}h^{-1}M_\odot$, of which the dominant component is dark matter. These are homes to galaxies, stars and eventually lives, which together hold the vast majority of the information that can be extracted from cosmological and astrophysical observations. In dark energy or modified gravity theories, the different cosmic expansion histories and gravitational laws between particles can have sizeable effects on how the clusters form and evolve. \citet{schmidt2009, cataneo2014},  based on this observation, have placed constraints on $f(R)$ gravity using cluster abundance data. In theories such as $f(R)$ gravity, massive and massless particles feel different strengths of gravity, thus allowing these theories to be constrained by comparing the so-called dynamical and lensing masses of clusters \citep{schmidt2010,zlk2011}. Combining with lensing observations, \citet{terukina2014,wilcox2015} obtained even stronger constraints on $f(R)$ gravity.

Given the abundant information associated with these rich objects, one expects that they will provide a wealth of other potential tests of new cosmological models. The arrival of the era of precision cosmology lends this perspective both more interest and more support. In this paper, we propose to utilise the observationally inferred mass fraction of hot X-ray emitting gas in galaxy clusters as a new test of gravity. 

The gas fraction in clusters, $f_{\rm gas}$, is a well established and understood observable in the standard cosmological model, which can be used to place strong constraints on background cosmology \citep[see, e.g.,][and following-up works, for some examples]{allen2004,allen2008}. The basic assumptions (or approximations) are (i) clusters are in hydrostatic equilibrium between thermal pressure and gravity and (ii) as the largest objects in the Universe, the cluster baryon fraction, dominantly contributed by gas, is a faithful representation of the cosmological average baryon fraction $\Omega_{\rm b}/\Omega_{\rm m}$, in which $\Omega_{\rm b}$ and $\Omega_{\rm m}$ are respectively the fractional mass density of baryons and all matter \citep{white1993,eke1998}. Using (i), one can find, from the observed X-ray temperature and surface brightness profiles, the mass profiles of baryons and all matter inside a cluster, and consequently $f_{\rm gas}(r)$ -- the profile of gas fraction. Combined with (ii), this can have a say about $\Omega_{\rm m}$ provided that $\Omega_{\rm b}$ is measured elsewhere (e.g., from big bang nucleosynthesis or the cosmic microwave background (CMB)). 

In modified gravity theories, the hydrostatic equilibrium inside clusters is changed as a result of the different law of gravity. Hence, a cluster can have a higher dynamical mass, with the same baryonic mass inside, leading to a $f_{\rm gas}^{\rm obs}$ that is lower than the true $f_{\rm gas}$ which is related to $\Omega_{\rm b}/\Omega_{\rm m}$. If a cosmologist wishes to infer $\Omega_{\rm b}/\Omega_{\rm m}$ from $f^{\rm obs}_{\rm gas}$ in a modified gravity universe, a correction has to be done to the end result, which can lead to inconsistency with other observational determinations of $\Omega_{\rm b}/\Omega_{\rm m}$, and therefore a constraint on the gravity theory. We shall demonstrate the potential constraints from this test using $f(R)$ gravity as an example.

This paper is organised as following: in Sect.~\ref{sect:fr_gravity} we briefly describe the $f(R)$ gravity theory and its equations which will be used in the discussion below. In Sect.~\ref{sect:gas_physics} we give a more detailed account of the physics related to the gas fraction test described above. Then we present a numerical example in Sect.~\ref{sect:numerical_examples}, which shows how current data of cluster gas fraction alone can give powerful constraints on gravity. We discuss the results and their implications in Sect.~\ref{sect:con}.

\section{The $f(R)$ gravity theory}

\label{sect:fr_gravity}

This section is devoted to a quick overview of $f(R)$ gravity. It will be kept brief and only include essential equations, given that there are already many papers in the literature covering this topic. 

%\subsection{The $f(R)$ gravity model}
%
%\label{subsect:fr}

$f(R)$ gravity \citep{cddatt2005} is a simple generalisation of GR, by replacing the Ricci scalar, $R$, in the Einstein-Hilbert action with an algebraic function $f(R)$
\begin{eqnarray}\label{eq:fr_action}
S = \int{\rm d}^4x\sqrt{-g}\left\{\frac{1}{2}\mpl^2\left[R+f(R)\right]+\mathcal{L}_{\rm m}\right\},
\end{eqnarray}
where $\mpl$ is the reduced Planck mass, $\mpl^{-2}=8\pi G$ with $G$ being Newton's constant, $g$ is the determinant of the metric $g_{\mu\nu}$ and $\mathcal{L}_{\rm m}$ is the Lagrangian density for (normal plus dark) matter fields. The model is defined by specifying the functional form of $f(R)$.

The action in Eq.~(\ref{eq:fr_action}) leads to a modified Einstein equation
\begin{eqnarray}\label{eq:fr_einstein}
G_{\mu\nu} + f_RR_{\mu\nu} -\left[\frac{1}{2}f-\Box f_R\right]g_{\mu\nu}-\nabla_\mu\nabla_\nu f_R = 8\pi GT^{\rm m}_{\mu\nu},
\end{eqnarray}
in which $G_{\mu\nu}$, $R_{\mu\nu}$ are respectively the Einstein and Ricci tensors, $f_R\equiv \rd f/\rd R$, $\nabla_{\mu}$ the covariant derivative compatible to the metric $g_{\mu\nu}$, $\Box\equiv\nabla^\alpha\nabla_\alpha$ and $T^{\rm m}_{\mu\nu}$ is the matter energy momentum tensor. Eq.~(\ref{eq:fr_einstein}) can be considered as the standard Einstein equation of GR with an extra scalar field, $f_R$, whose dynamics is governed by
\begin{eqnarray}
\Box f_R = \frac{1}{3}\left(R-f_RR+2f+8\pi G\rho_{\rm m}\right),
\end{eqnarray}
where $\rho_{\rm m}$ is the mass density of baryons and dark matter. As we are interested in late times, photons and neutrinos will be neglected.

%Assuming that the background Universe is described by the flat Friedmann-Robertson-Walker (FRW) metric, the line element in the perturbed Universe is written as
%\begin{eqnarray}
%\rd s^2 = a^2(\eta)\left[(1+2\Phi)\rd\eta^2 - (1-2\Psi)\rd x^i\rd x_i\right],
%\end{eqnarray}
%in which $\eta$ and $x^i$ are respectively the conformal time and comoving coordinates, $\Phi(\eta,{\bf x})$ and $\Psi(\eta,{\bf x})$ are the Newtonian potential and perturbation to the spatial curvature, and are functions of both time $\eta$ and space ${\bf x}$; $a$ denotes the scale factor of the Universe and $a=1$ today.

On scales well inside the Hubble radius, and for the models to be considered, it is safe to work with the quasi-static approximation \citep{bose2015}, in which the scalar equation becomes
\begin{eqnarray}\label{eq:fr_eqn_static}
\vec{\nabla}^2f_R = -\frac{1}{3}a^2\left[R(f_R)-\bar{R} + 8\pi G\left(\rho_{\rm m}-\bar{\rho}_{\rm m}\right)\right],
\end{eqnarray}
where $\vec{\nabla}$ denotes the three dimensional gradient, $a$ is the scale factor, and an overbar takes the background value of a quantity. Notice that $R$ can be expressed as a function of $f_R$ by inverting $f_R(R)$.

Similarly, the modified Poisson equation in this limit reads as
\begin{eqnarray}\label{eq:poisson_static}
\vec{\nabla}^2\Phi = \frac{16\pi G}{3}a^2\left(\rho_{\rm m}-\bar{\rho}_{\rm m}\right) + \frac{1}{6}a^2\left[R\left(f_R\right)-\bar{R}\right],
\end{eqnarray}
where $\Phi$ is the Newtonian potential.

Eq.~(\ref{eq:fr_eqn_static}) implies two limits of the behaviour of $f(R)$ gravity:

(i) When $f_R$ is small, or more accurately, when $|f_R|\ll|\Phi|$, it recovers the well-known GR solution $R=8\pi G\rho_{\rm m}$, and so Eq.~(\ref{eq:poisson_static}) reduces to GR as well. This is the {\it chameleon} \citep{chameleon} regime which any viable $f(R)$ model must be in to pass the stringent solar system and terrestrial tests of gravity. 

(ii) When $|f_R|\sim\mathcal{O}(|\Phi|)$, the second term on the rhs of Eq.~(\ref{eq:poisson_static}) is negligible compared with the first term, so that we have a gravity that is $1/3$ stronger than in GR. This is ususally known as the {\it non-chameleon}, or {\it unscreened}, regime.

It is evident that the unscreened regime mostly happens where $\Phi$ is shallow, or in extensive regions of low density. On large scales, matter density is close to the cosmological average, and so the total gravity in $f(R)$ gravity is enhanced within scales comparable to the Compton wavelength of the scalar field $f_R$ (which in most models of interest is in the range $\mathcal{O}(1\sim10)h^{-1}$Mpc). This naturally leads to an enhanced large-scale structure formation, and features such as over-abundant and more massive galaxy clusters - a topic which has been extensively studied previously. This will also be the topic that we focus on in this paper.

\section{Cluster gas fraction}

\label{sect:gas_physics}

Galaxy clusters are the largest bound objects in the Universe, whose masses are dominated by the dark matter component, with the baryonic masses dominated by X-ray emitting intracluster gas, which is heated to temperatures of the order of keV during virialisation. It is the mass fraction of this gas component that we will employ to test the theory of gravity here.

%They are a rich source of observational data, from which constraints on cosmological models and parameters can be and have been drawn. 

In this section, we shall first give a brief overview of how the baryon fraction can be estimated observationally, and how it can be used to constrain cosmological models and their parameters. Then, we will discuss how this process might be affected if the underlying theory of gravity is modified. For simplicity, we shall neglect other baryonic components than the intracluster gas in our analysis %, and use gas fraction and baryon fraction interchangeably 
unless otherwise stated. 

\subsection{The standard $\Lambda$CDM model}

\label{sect:standard_model}

In the standard cosmological scenario, halo density profiles can be universally described by the \citet[][NFW]{nfw} fitting formula, which is often expressed as
\begin{eqnarray}
\rho(r) = \frac{\rho_{\rm s}}{(r/r_{\rm s})\left(1+r/r_{\rm s}\right)^2},
\end{eqnarray}
in which $\rho(r)$ is the halo mass density as a function of the distance, $r$, from the halo's centre, $\rho_{\rm s}$ is a characteristic density and $r_{\rm s}$ is the scale radius. We shall assume that the halo is spherically symmetric and well relaxed throughout the analysis, unless otherwise stated.

The mass of the halo can be obtained by integrating the NFW profile from $r=0$ to $r=R_{\Delta}$, in which $R_{\Delta}$ is the edge of the halo and is defined as the radius within which the average mass density is $\Delta\times\rho_{\rm crit}(z)$, with $\rho_{\rm crit}(z)\equiv3H(z)^2/8\pi G$ the critical density at the redshift $z$ when the halo is identified. This leads to
\begin{eqnarray}
M_{\rm halo} = 4\pi\rho_{\rm s}r^3_{\rm s}\left[\ln\left(1+R_\Delta/r_{\rm s}\right)-\frac{R_\Delta/r_{\rm s}}{1+R_\Delta/r_{\rm s}}\right].
\end{eqnarray}

Observationally, the total and baryonic masses of a cluster %that is assumed to have a NFW profile 
can be obtained by measuring its X-ray surface brightness profile, and the temperature profile of its X-ray gas. For a dynamically relaxed system that consists of dark matter and baryonic gas, a hydrostatic equilibrium can be achieved, which satisfies the following equation
\begin{eqnarray}\label{eq:hydrostatic}
\frac{1}{\rho_{\rm gas}(r)}\frac{{\rm d}}{{\rm d}r}P_{\rm gas}(r) = -\frac{GM_{\rm tot}(<r)}{r^2},
\end{eqnarray}
in which $M_{\rm tot}(<r)$ is the total mass of dark matter and gas within radius $r$, and $\rho_{\rm gas}(r)$, $P_{\rm gas}(r)$ are respectively the density and pressure of the gas at $r$. For simplicity, we neglect  non-thermal pressure in our discussion (the effects of non-thermal pressure, however, are taken into account in the error budget when modelling the relation between $f_{\rm gas}$ and $\Omega_{\rm b}/\Omega_{\rm m}$, cf.~Eq.~(\ref{eq:allen_model}) below). 

For an ideal thermal gas, its pressure and density are related to its temperature, $T_{\rm gas}$, as
\begin{eqnarray}\label{eq:eos}
P_{\rm gas} = kn_{\rm gas}T_{\rm gas} = \frac{k}{\mu m_{\rm p}}\rho_{\rm gas}T_{\rm gas},
\end{eqnarray}
where $k$ is the Boltzmann constant, and the mass and number densities of the gas particles are connected by $\rho_{\rm gas}=\mu m_{\rm p}n_{\rm gas}$, where $m_{\rm p}$ is the proton mass and $\mu$ the mean molecular weight. Applying Eqs.~(\ref{eq:hydrostatic}, \ref{eq:eos}), we obtain
\begin{eqnarray}\label{eq:hydrostatic_spherical}
\frac{GM(<r)}{r} = -\frac{kT_{\rm gas}(r)}{\mu m_{\rm p}}\left[\frac{{\rm d}\ln\rho_{\rm gas}(r)}{{\rm d}\ln r}+\frac{{\rm d}\ln T_{\rm gas}(r)}{{\rm d}\ln r}\right],
\end{eqnarray}
which, if evaluated at the halo edge ($r=R_\Delta$), gives
\begin{eqnarray}\label{eq:halo_mass}
\frac{GM_{\rm halo}}{R_\Delta} = -\frac{kT_{\rm gas}(R_\Delta)}{\mu m_{\rm p}}\left[\frac{{\rm d}\ln\rho_{\rm gas}}{{\rm d}\ln r}+\frac{{\rm d}\ln T_{\rm gas}}{{\rm d}\ln r}\right]_{r=R_\Delta}.
\end{eqnarray}

In the mean time, the X-ray emission of galaxy clusters is produced mainly by thermal bremsstrahlung radiation, leading to a X-ray surface brightness profile that also depends on the gas density and temperature profiles, $\rho_{\rm gas}(r)$ and $T_{\rm gas}(r)$. In terms of gas density, the gas mass can be expressed as
\begin{eqnarray}\label{eq:gas_mass}
M_{\rm gas}(<r) = \int^r_04\pi r'^2{\rm d}r'\rho_{\rm gas}(r') \propto \rho_{\rm gas}(0)r_0^3,
\end{eqnarray}
where $\rho_{\rm gas}(0)$ is the gas density at the cluster centre, and we have assumed that $\rho_{\rm gas}(r)=\rho_{\rm gas}(0)g(r/r_0)$ with $g(x)$ some function describing the profile and $r_0$ a characteristic scale. In the isothermal $\beta$ model, for example, we have 
\begin{eqnarray}
\rho_{\rm gas}(r) = \rho_{\rm gas}(0)\left[1+\left(\frac{r}{r_0}\right)^2\right]^{-3\beta/2},
\end{eqnarray}
where $\beta$ is a dimensionless constant. 

The bolometric luminosity is given by \citep[e.g.,][]{sasaki1996}
\begin{eqnarray}\label{eq:gas_luminosity}
L_X(<r) \propto \int^r_04\pi r'^2{\rm d}r'T^{1/2}_{\rm gas}(r')\rho_{\rm gas}^2(r') \propto \rho_{\rm gas}^2(0)r_0^3,
\end{eqnarray}
where we have neglected the proportionality coefficients which are irrelevant here.

Combining Eqs.~(\ref{eq:gas_mass}, \ref{eq:gas_luminosity}), we have 
\begin{eqnarray}\label{eq:temp1}
M_{\rm gas} \propto L_X^{1/2}r_0^{3/2} \propto d_{\rm L}f^{1/2}_X\Theta_0^{3/2}d_{\rm A}^{3/2} \propto (1+z)^2d^{5/2}_{\rm A},
\end{eqnarray}
where $d_{\rm L}$ and $d_{\rm A}$ are respectively the luminosity and angular diameter distances and are related by $d_{\rm L}=(1+z)^2d_{\rm A}$. Here we have used $L_X=4\pi d_{\rm L}^2f_X$ where $f_X$ is the X-ray flux, and $r_0=d_{\rm A}\Theta_0$, in which $\Theta_0$ denotes the angle spanned by $r_0$ at redshift $z$. Note that $f_X$ (or equivalently the surface brightness) and $\Theta_0$ are the observed quantities in this description. In real observations, one has the surface brightness and temperature profiles, or equivalently $L_X(<r)$ and $T_{\rm gas}(r)$, using which Eqs.~(\ref{eq:hydrostatic_spherical}, \ref{eq:gas_luminosity}) can be solved simultaneously to find $M_{\rm halo}(<r)$ and $M_{\rm gas}(<r)$ ({note that the innermost regions of clusters are often excluded due to complicated processes such as cooling flow. For example, the study of \citet{allen2008} uses only the data within $0.7\sim1.2 R_{2500}$ to measure $f_{\rm gas}$}). These provide the necessary information to find
\begin{eqnarray}\label{eq:xxx}
f_{\rm gas}(r) = \frac{M_{\rm gas}(<r)}{M_{\rm halo}(<r)}.
\end{eqnarray}
Often in observations, people quote the value of $f_{\rm gas}$ at $r=R_{2500}$ as the cluster gas fraction.  We note in passing that here $M_{\rm halo}$ enters the picture only through its gravitational effect on gas particles, as this fact is important for the discussion of modified gravity below.

To see how this can be used to constrain background cosmology, we note that, once the right-hand side of Eq.~(\ref{eq:halo_mass}) is known by observations, %as well as the apparent angular scale $\Theta_\Delta=R_\Delta/d_{\rm A}$ spanned by $R_\Delta$, 
we also have a fixed numerical value of the left-hand side and hence have the relation
\begin{eqnarray}\label{eq:temp2}
M_{\rm halo} \propto R_\Delta \propto d_{\rm A}.
\end{eqnarray}
Eqs.~(\ref{eq:temp1}, \ref{eq:temp2}) imply that the {measured} cluster gas fraction depends on redshift and angular diameter distance in the following specific way:
\begin{eqnarray}\label{eq:apparent_gas_fraction}
f_{\rm gas}(z) = \frac{M_{\rm gas}}{M_{\rm halo}} \propto (1+z)^2d_{\rm A}^{3/2}.
\end{eqnarray}
As mentioned earlier, we expect the cluster gas fraction to be a reasonably fair sample of the mean cosmological baryon fraction and is therefore roughly independent of $z$ for massive clusters at low $z$. This is a reasonable assumption which should hold regardless of the cosmological model/parameters. The measurement of the apparent gas fraction, as described above, involves the angular diameter distance $d_{\rm A}$ and is indeed dependent on both the cosmological model and its parameters. Therefore, with incorrect cosmological models or parameters, the constancy of the true gas fraction (hereafter $f_{\rm gas}^{\ast}$) is not guaranteed to be reflected in the observed value $f^{\rm obs}_{\rm gas}(z)$. This provides a powerful test \citep{allen2004} of {\it background} cosmology and can be used to constrain cosmological parameters.

Were galaxy clusters perfectly fair samples of the average matter components in the Universe, their baryon fraction would just be $\Omega_{\rm b}/\Omega_{\rm m}$. The true situation, however, is more complicated. To take the complexities into account, \citet{allen2008}, improving on the earlier work of \citet{allen2004}, propose the following model of the relation between cluster gas fractions and $\Omega_{\rm b}/\Omega_{\rm m}$,
\begin{eqnarray}\label{eq:allen_model}
f^\ast_{\rm gas} = \frac{K\gamma b(z)}{1+s(z)}\frac{\Omega_{\rm b}}{\Omega_{\rm m}},
\end{eqnarray}
in which:

$\bullet$ $K$ is a constant accounting for systematic effects such as the calibration of instrument and X-ray modelling -- it is assumed to be $K=1.0\pm0.1$ in \citet{allen2008};

$\bullet$ $\gamma$ models the non-thermal pressure support in galaxy clusters which can cause a bias in the estimate of $f^\ast_{\rm gas}$ of about $9$\%; 

$\bullet$ $b(z)\equiv b_0(1+\alpha_{\rm b}z)$ is the so-called 'depletion' factor which is inspired by the observation that the baryon fraction at $R_{2500}$ in non-radiative simulations \citep{eke1998} is actually smaller than $\Omega_{\rm b}/\Omega_{\rm m}$, with $b_0=0.83\pm0.04$ and $\alpha_{\rm b}$ small indicating a weak redshift evolution below $z=1$;

$\bullet$ $s(z)\equiv s_0(1+\alpha_{\rm s}z)$ accounts for the fact that a small fraction of baryons can be in the form of stars, with $s_0=0.16\pm0.048$ and $-0.2<\alpha_{\rm s}<0.2$ describing its redshift evolution. 

%There is also a correction factor $A$ in the original \citet{allen2008} model describing the impact of a different background cosmology on the apparent angular sizes of clusters. For our discussion below, we will discard this factor because the background cosmology of all models considered here is practically the same.

The model in Eq.~(\ref{eq:allen_model}) indeed has a weak redshift dependency. However, any additional dependency from the observed $f_{\rm gas}$ would imply that one is using the wrong background cosmology to extract data, cf.~Eq.~(\ref{eq:apparent_gas_fraction}). This is, to be clear, in the framework of standard GR.

\begin{figure*}
 \includegraphics[width=180mm]{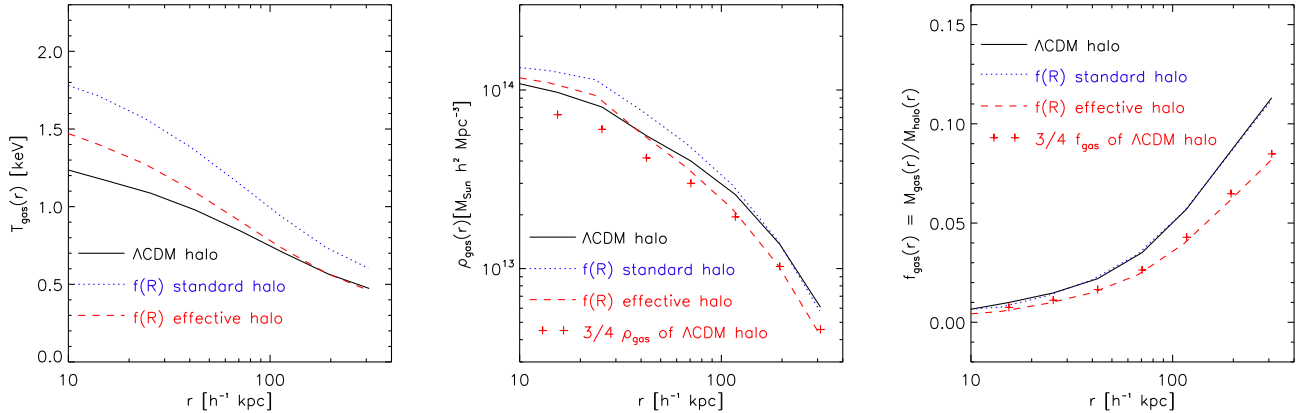}
\caption{(Colour online) {\it Left panel}: the halo gas temperature profiles from the non-radiative hydrodynamical simulations of \citet{he2015b} for the standard $\Lambda$CDM model (black solid line) and $f(R)$ gravity with $|\bar{f}_{R0}|=10^{-5}$ (blue dotted line for standard haloes and red dashed line for effective haloes. The same line styles are used for the other two panels). {\it Middle panel}: The same as the left panel, but for the gas density profiles. The red crosses are a rescaling of the $\Lambda$CDM curve by $3/4$ to take into account the fact that for a $\Lambda$CDM halo and a $f(R)$ effective halo of the same mass, their true masses differ by $1/3$. {\it Right panel}: The halo gas fraction profiles from the same simulations; the red crosses are the $\Lambda$CDM result scaled by $3/4$, which is almost identical to the red dashed line, cf.~Eq.~(\ref{eq:observer_wrong}). The profiles shown here are stacked results of the haloes in the mass range $10^{13}\sim10^{13.4}h^{-1}M_\odot$, for illustration purpose; we've checked that results from other halo mass bins follow the same trends.}
\label{fig:simulation}
\end{figure*}

\subsection{Modified gravity scenarios}

In many modified gravity theories, including $f(R)$ gravity, the way in which the trajectories of massive test bodies -- e.g., galaxies, stars and gas particles -- respond to the underlying matter distribution is different. This change of the dynamics of test bodies is sometimes described as the change of the {\it dynamical mass} of matter. Massless particles, such as photons, could behave differently: in some theories, such as the Galileon model \citep{galileon1, galileon2} and the K-mouflage model \citep{kmouflage1, kmouflage2}, photons can also feel a different mass of matter, but in other models, for example $f(R)$ gravity and the \citet[][DGP]{dgp} model, photon trajectories depend on the matter distribution in essentially the same way as in GR, since the conformal coupling does not affect geodesics of massless particles. For distinction, the mass felt by photons is usually called the {\it lensing mass}. The differences in the dynamical and lensing masses of galaxy clusters have been used to constrain $f(R)$ gravity in, e.g., \citet[][]{terukina2014,wilcox2015}

As mentioned above, gas particles, like dark matter particles and galaxies, do feel the dynamical mass of a cluster. In $f(R)$ gravity, the same cluster can have a dynamical mass $4/3$ times its value in GR. This maximum enhance factor of 1/3, however, is not necessarily realised in all clusters, because of the chameleon screening \citep{chameleon}. The screening helps to reduce the difference between the dynamical and lensing masses, especially for more massive clusters. Consequently, constraints relying on the dynamical masses of clusters are in general weaker than those coming from astrophysical considerations. Nevertheless, they have cleaner physics than that of astrophysical observables -- which can often depend on whether the considered astrophysical system lives inside a screened cluster -- and are amongst the tightest constraints obtained using cosmological data \citep{terukina2014}. 

Evidently, without a reliable measurement of the lensing mass, it is difficult to tell whether one observes a cluster of true mass $M$ in GR, or one with a smaller mass in $f(R)$ gravity, since both have the same dynamical masses. This has motivated \citet{he2015a} to propose the concept of {\it effective haloes}. Briefly speaking, the idea is to redefine the right-hand side of the modified Poisson equation in $f(R)$ gravity, Eq.~(\ref{eq:poisson_static}), 
%\begin{eqnarray}\label{eq:poisson}
%\vec{\nabla}^2\Phi = \frac{16\pi G}{3}a^2\delta\rho_{\rm m}+ \frac{1}{6}a^2\left[R\left(f_R\right)-\bar{R}\right],
%\end{eqnarray}
so that it can be rewritten as
\begin{eqnarray}\label{eq:poisson_eff}
\vec{\nabla}^2\Phi = 4\pi Ga^2\delta\rho_{\rm m,eff}.
\end{eqnarray}
In the above, $\Phi$ is the Newtonian potential that determines the dynamics of massive bodies and $\delta\rho_{\rm m}\equiv\rho_{\rm m}-\bar{\rho}_{\rm m}$ is the density perturbation of (dark plus baryonic) matter. $\delta\rho_{\rm m,eff}$ is the {\it effective density field}, with which the Poisson equation takes exactly the same form as in GR [cf.~Eq.~(\ref{eq:poisson_eff})]. In this way, the complicated new physics in $f(R)$ gravity is absorbed into $\rho_{\rm m,eff}$, and with that solved (e.g., in numerical simulations) one can in principle proceed assuming GR as the true theory of gravity. \citet{he2015b}, for example, show with hydrodynamical simulations that cluster gas temperatures depend only on the masses of the corresponding effective haloes, and that with certain rescaling depending on the effective halo mass the scaling relations -- such as the $L_X$-$M$ relation with $L_X$ the X-ray luminosity -- in $f(R)$ gravity can be derived reliably using existing knowledge of GR. 

In the left panel of Fig.~\ref{fig:simulation}, we present the gas temperature profiles for standard haloes in GR and effective haloes in $f(R)$ gravity, both in the mass bin $10^{13}\sim10^{13.4}h^{-1}M_\odot$. Though there are differences in the inner regions -- which could be due to different halo density profiles or screening -- we notice that beyond $\sim100h^{-1}$kpc the two agree very well. \citet{he2015b} find that the average gas temperatures in the two also show very good agreement, and indeed the temperature-mass relation is barely distinguishable in the two models, provided that effective haloes are used in $f(R)$ gravity.

The fact that the cluster gas temperatures depend on the mass of the effective haloes is as expected, since for relaxed systems the virial temperature depends on the Newtonian potential, which does not distinguish between standard (GR) and effective ($f(R)$ gravity) density fields. For a polytropic gas with an equation of state $P_{\rm gas}\propto\rho_{\rm gas}^\Gamma$ in which the constant $\Gamma\geq1$, the hydrostatic equation implies that \citep[see, e.g.,][]{mvw-book} the temperature can be analytically expressed as a function of the potential $\Phi$.

Let us now consider two haloes, one identified in the standard dark matter field in a model with GR as the gravity theory, another from the effective density field in $f(R)$ gravity. The profiles of the two haloes are the same so that Eq.~(\ref{eq:poisson_eff}) sees no difference in them. Since gravity only enters the picture through Eq.~(\ref{eq:hydrostatic_spherical}), we make the following two observations/predictions:

$\bullet$ the gas temperature profiles are the same in these two haloes;

$\bullet$ the two sides of the spherical hydrostatic equation, Eq.~(\ref{eq:hydrostatic_spherical}), are the same for the two haloes.

\begin{figure*}
 \includegraphics[width=182mm]{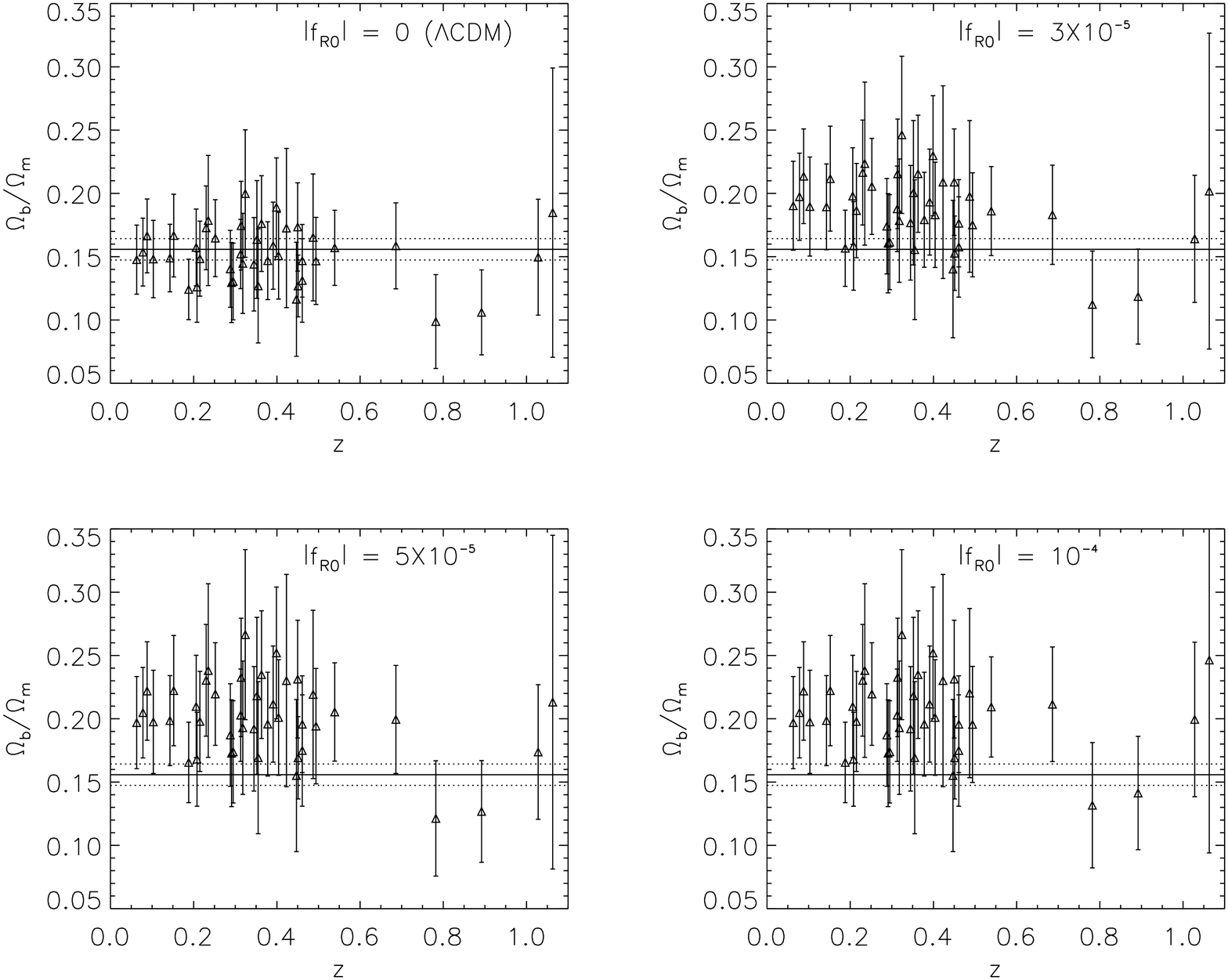}
\caption{The inferred $\Omega_{\rm b}/\Omega_{\rm m}$ (triangles with 1$\sigma$ error bars) using the $f^{\rm obs}_{\rm gas}$ for the 42 clusters from \citet{allen2008}, as a function of the cluster redshift. 4 cases are shown, with different underlying  models of gravity: GR (upper left), and Hu-Sawicki $n=1$ $f(R)$ model with $|\bar{f}_{R0}|=3\times10^{-5}$ (upper right), $5\times10^{-5}$ (lower left), $10^{-4}$ (lower right). The horizontal solid and dotted lines are respectively the mean and 1$\sigma$ range of $\Omega_{\rm b}/\Omega_{\rm m}$ from Planck CMB data. A good match between $f_{\rm gas}$ and CMB for the GR case, and progressively worse matches for the $f(R)$ models, can be seen by a quick inspection by eye. For simplicity, all haloes are assumed to have a mass of $7.5\times10^{14}h^{-1}M_\odot$ and a concentration of $3.3$ when determining the effect of the chameleon screening.}
\label{fig:obom}
\end{figure*}

Since $T_{\rm gas}(r)$ and ${\rm d}\ln T_{\rm gas}(r)/{\rm d}\ln r$ in Eq.~(\ref{eq:hydrostatic_spherical}) are the same, we conclude that ${\rm d}\ln\rho_{\rm gas}(r)/{\rm d}\ln r$ is also the same in the haloes. This, however, does not necessarily mean that the haloes have identical gas density profiles, because we can rescale $\rho_{\rm gas}$ by a constant factor without changing ${\rm d}\ln\rho_{\rm gas}(r)/{\rm d}\ln r$. To confirm this, in the middle panel of Fig.~\ref{fig:simulation} we compare the gas density profile of $f(R)$ effective haloes with that of $\Lambda$CDM haloes of the same mass, and find that the two show a constant shift by 1/4 beyond $r\sim100h^{-1}$ kpc. Note that the simulated haloes in the plot do not have perfectly identical total -- standard or effective -- mass profiles, which is why in the middle panel of Fig.~\ref{fig:simulation} the red crosses and red dashed line do not agree on scales below $\sim100h^{-1}$kpc. However, as mentioned above Eq.~(\ref{eq:xxx}), in real observations, such innermost regions are not used in the determination of $f_{\rm gas}$ anyway.

As a result, to obtain the gas density profile, we need further, independent, information to fix its normalisation -- as opposed to its shape -- which brings us back to the measurements of cluster X-ray surface brightness. Inspecting the equation for the cluster X-ray luminosity, Eq.~(\ref{eq:gas_luminosity}), we notice that the luminosity density (i.e., the integrand) depends on (i) the {\it physical} gas density $\rho_{\rm gas}$, and (ii) the gas temperature $T_{\rm gas}$ which, as we have seen above, depends on the total mass of the {\it effective} halo. Consequently, should the physical gas densities be the same for $f(R)$ effective and $\Lambda$CDM haloes of the same mass, there would be no difference in their X-ray surface brightness profiles.

%Therefore, if effective haloes, instead of standard haloes, are used for $f(R)$ gravity, then $L_X$ can be analysed similarly as, and compared directly to, $\Lambda$CDM results \cite{he2015b}.

However, despite the standard (GR) and effective ($f(R)$ gravity) haloes above having identical gas temperature and halo mass, their {\it actual} (physical, or lensing) masses are different, and it is important to remember that $\Omega_{\rm m}$ characterises the amount of the {\it actual} mass in the Universe. If, as we have assumed so far, the gas fraction in clusters is a fair sample of the cosmological value, it would be the ratio of the gas mass and {\it actual} halo's mass that satisfies Eq.~(\ref{eq:allen_model}). Gas fractions inferred observationally, in the way described in the previous subsection, are in fact the ratio of the gas mass and that of the {\it effective} halo. If we denote the ratio of the effective and actual masses of a halo by $\eta$, then $1\leq\eta\leq4/3$, depending on the actual mass and environment of the halo, its redshift, as well as the $f(R)$ model parameters\footnote{More accurately speaking, effective haloes are identified from the effective density field, $\rho_{\rm m,eff}$ in Eq.~(\ref{eq:poisson_eff}), and standard haloes are identified from the physical density field $\rho_{\rm m}$. They do not necessarily share the same physical particles. Here, for simplicity, when talking about the effective and actual masses of some halo, we mean the masses of the effective and standard haloes that would be considered as matched haloes in the two catalogues.}. Here, as we are interested in the most massive clusters, with $M_{\rm halo}\sim10^{14}-10^{15}h^{-1}M_{\odot}$, we can for simplicity neglect the impacts of the halo's environment, so that $\eta$ mainly depends on $z$, i.e., $\eta=\eta(z)$, for a given halo mass.

Consider the extreme case in which $\eta=4/3$ as example. The apparent $f_{\rm gas}$ inferred from X-ray cluster observations would be 
\begin{eqnarray}\label{eq:observer_wrong}
f^{\rm obs}_{\rm gas} = \frac{M_{\rm gas}}{M_{\rm halo, eff}} = \frac{3}{4}\frac{M_{\rm gas}}{M_{\rm halo,actual}} = \frac{3}{4}f^\ast_{\rm gas}.
\end{eqnarray}
This is also confirmed by hydro simulations, as shown in the right panel of Fig.~\ref{fig:simulation}. In that plot, we see that if haloes are defined using their actual masses in $f(R)$ gravity, then they share the same $f_{\rm gas}$ as $\Lambda$CDM haloes of the same mass (black solid vs. blue dotted lines). For effective $f(R)$ haloes, on the other hand, their $f_{\rm gas}$ profiles are a constant downward shift by $1/4$ from the results of $\Lambda$CDM haloes of the same masses (red dashed line vs.~red crosses), which is just what Eq.~(\ref{eq:observer_wrong}) predicts.

It is worthwhile to pause for a moment and try to understand the physics behind the behaviour of Fig.~\ref{fig:simulation}. It may seem surprising that, although the gas density profiles in $\Lambda$CDM and $f(R)$ standard haloes are significantly different within $r\sim100h^{-1}$kpc, their gas fraction profiles are very close to each other. This suggests that the two also have different dark matter (or total) mass profiles to cancel the differences in $\rho_{\rm gas}(r)$. In other words, $\rho_{\rm gas}(r)$ follows $\rho_{\rm DM}(r)$, which is the {\it physical} dark matter mass density, in the same way in $\Lambda$CDM and $f(R)$ standard haloes, even though they have different mass and even more different potential profiles. To understand this, we note that in Eq.~(\ref{eq:hydrostatic_spherical}) the $G$ on the left-hand side and $T_{\rm gas}$ on the right-hand side are both modified in $f(R)$ standard haloes. Indeed, if one assumes hydrostatic equilibrium and $P_{\rm gas}\propto\rho^\Gamma_{\rm gas}$, then the gas density profile can be written as \citep{komatsu2001}:
\begin{eqnarray}\label{eq:xxx}
\frac{\rho^{\Gamma-1}_{\rm gas}(r)}{\rho^{\Gamma-1}_{\rm gas}(0)} = 1- \frac{\Gamma-1}{\Gamma}\frac{G\mu m_{\rm p}M_{\rm halo}}{R_{\rm vir}kT_{\rm gas}(0)}\frac{c}{m(c)}\int^{\frac{r}{r_{\rm s}}}_0\frac{m(x)}{x^2}{\rm d}x,
\end{eqnarray}
in which $m(r/r_{\rm s})\equiv M(<r)/4\pi\rho_{\rm s}r^3_{\rm s}$. Eq.~(\ref{eq:xxx}) is derived under the assumption of self-similarity of gas density profiles, but the key point therein, that the modified gravity effects on $G$ and $T_{\rm gas}$ can be cancelled out, is not affected by this assumption. If this cancellation happens, then the gas density profile is determined by the total mass profile under the assumption of hydrostatic equilibrium, regardless of the theory of gravity\footnote{Note in Eq.~(\ref{eq:xxx}) it is $\rho_{\rm gas}(r)/\rho_{\rm gas}(0)$ that is determined by $m(r/r_{\rm s})$. The normalisation of $\rho_{\rm gas}$ will then be fixed by the total gas fraction inside the halo.}. 

In clusters, gas is heated by accretion shocks during the assembly of the halo, a process which involves the conversions of energy from gravitational to kinetic (that of the cold accreted gas) and then to thermal (via shocks). Assuming a complete thermalisation, the post-shock gas temperature is proportional to $v_{\rm infall}^2$, with $v_{\rm infall}$ the infall speed of the accreted gas \citep[e.g.,][]{mvw-book}. Consequently, energy conservation implies that the final gas temperatures in the central regions will be affected in the same way as $v_{\rm infall}^2$ of the cold gas and hence $G$ in modified gravity.
Of course, this is only an approximation, and the cancellation of the effects of modified gravity on $G$ and $T_{\rm gas}$ depend on various factors including the screening and formation history of a cluster, which is not expected to be complete. However, Fig.~\ref{fig:simulation} suggests that it works pretty well for the haloes we use here. We checked explicitly that it works slightly less well for more massive haloes, for which the agreement between the $f_{\rm gas}(r)$ in $\Lambda$CDM and $f(R)$ standard haloes is slightly less perfect -- this may be because those haloes became unscreened only very recently.

The argument above in theory also applies to effective haloes, for which $G$ is the same as in GR, but the effects of modified gravity are incorporated in $M_{\rm halo}$. However, in the effective halo case the normalisation is different because of the different total gas fraction (see footnote 2) -- although the shape is the same -- hence the nearly constant rescaling of the red dashed curve compared with the black solid one in Fig.~\ref{fig:simulation}.

Coming back to the discussion prior to the previous three paragraphs, our result suggests two possible tests of $f(R)$ gravity:

(i) If an observer actually lives in a universe shaped by $f(R)$ gravity, then the true cluster gas fraction is given by $f^{\ast}_{\rm gas}=\eta f^{\rm obs}_{\rm gas}$. Assuming that Eq.~(\ref{eq:allen_model}) still holds for $f^\ast_{\rm gas}$, the observer will need to do the following transformation to get the true $\Omega_{\rm b}/\Omega_{\rm m}$:
\begin{eqnarray}
f^{\rm obs}_{\rm gas} = \frac{1}{\eta}\frac{K\gamma b(z)}{1+s(z)}\left[\frac{\Omega_{\rm b}}{\Omega_{\rm m}}\right]_{\rm true} \Rightarrow \left[\frac{\Omega_{\rm b}}{\Omega_{\rm m}}\right]_{\rm true} = \eta\left[\frac{\Omega_{\rm b}}{\Omega_{\rm m}}\right]_{\rm obs}.
\end{eqnarray} 
As $(\Omega_{\rm b}/\Omega_{\rm m})_{\rm obs}$ depends only on the actual observational data, the observer will obtain the same value as an observer in a standard GR universe would do. The resulting $(\Omega_{\rm b}/\Omega_{\rm m})_{\rm true}$ might then be too large to be compatible with other constraints, such as the one from the CMB.

(ii) Alternatively, if one takes the $\Omega_{\rm b}/\Omega_{\rm m}$ measured by other probes as the true value and starts from there, then Eq.~(\ref{eq:observer_wrong}) implies that the observed cluster gas fraction $f^{\rm obs}_{\rm gas}$ will be smaller than what the $\Lambda$CDM model and simulations predict. Because of the time dependence of $\eta(z)$ (see above), if the $f(R)$ model parameters happen to take the values for $\eta$ to evolve from 1 to $4/3$ between $z=1$ and the present for the clusters of interest, there may also be an apparent decrease of $f^{\rm obs}_{\rm gas}(z)$ as $z$ decreases, by a maximum of $25\%$.

\section{Numerical Examples}

\label{sect:numerical_examples}

In this section, we use a simplified example to illustrate the power of the cluster gas fraction test proposed above. For this, we will use the gas fraction data of the 42 clusters studied by \citet[][Table 3]{allen2008}. As described above, these $f_{\rm gas}$ data are obtained by fitting the gas temperature and X-ray surface brightness profiles of these clusters simultaneously, assuming NFW profiles for the total mass in clusters. We have also found that, in the context of $f(R)$ gravity, as long as we use effective haloes, the dynamics of gas particles can be calculated using standard gravity theory. Therefore, in this work {\it we can directly take the data of} \citet{allen2008} {\it as $f^{\rm obs}_{\rm gas}$}, bearing in mind that the cluster mass inferred therein would be the effective mass and therefore $f^{\rm obs}_{\rm gas}$ can be different from $f^{\ast}_{\rm gas}$ for unscreened clusters, cf.~Eq.~(\ref{eq:observer_wrong}). 

To obtain an estimation of the mean and standard deviation of $\left(\Omega_{\rm b}/\Omega_{\rm m}\right)_{\rm obs} = \frac{1+s(z)}{K\gamma b(z)}f^{\rm obs}_{\rm gas}$ from each cluster, random samples of size $10^5$ are drawn for each parameter or data: $K, \gamma, b_0, \alpha_{\rm b}, s_0, \alpha_{\rm s}$ and $f^{\rm obs}_{\rm gas}$. Of these, $f^{\rm obs}_{\rm gas}$ is taken, for a given cluster, from Table 3 of \citet{allen2008}, and is assumed to satisfy a Gaussian distribution with mean and standard deviation given by \citet{allen2008}. The other parameters and their distributions are shown in Table~\ref{tab:parameters}. We therefore obtain $10^5$ realisations of $\left(\Omega_{\rm b}/\Omega_{\rm m}\right)_{\rm obs}$, from which its mean and standard deviation can be calculated. This procedure is repeated for all 42 clusters.

\begin{table}
\caption{The assumed ranges and distributions of the model parameters in Eq.~(\ref{eq:allen_model}). For more details the readers are referred to Sect.~\ref{sect:standard_model} or \citet{allen2008}.}
\begin{tabular}{@{}llll}
\hline\hline
param & physical effect described & mean $\pm$ stddev & prior \\
\hline
$K$ & overall calibration & $1.000\pm0.100$ & Gaussian \\
$\gamma$ & non-thermal pressure & $1.050\pm0.050$ & Uniform \\
$b_0$ & gas bias: normalisation & $0.825\pm0.175$ & Uniform \\
$\alpha_{\rm b}$ & gas bias: evolution & $0.000\pm0.100$ & Uniform \\
$s_0$ & stellar fraction: normalisation & $0.160\pm0.048$ & Gaussian \\
$\alpha_{\rm s}$ & stellar fraction: evolution & $0.000\pm0.200$ & Uniform \\
\hline
\end{tabular}
\label{tab:parameters}
\end{table}

The estimation of the effects of modified gravity, i.e., the factor $\eta(z)$, is more complicated, since it depends on the cluster mass, density profile, environment, redshift, as well as the $f(R)$ parameters. Because the main purpose of this paper is to illustrate the basic idea, we shall leave a full analysis using real cluster data for future work, and instead adopt a simplified modelling. The cluster masses are assumed to be the same, with $M_{\rm halo}=7.5\times10^{14}h^{-1}M_\odot$, for all 42 clusters, since this is a typical value for massive X-ray clusters. The cluster's radius ($R_{200}$ or $R_{\rm vir}$) is taken to be $1.5h^{-1}$Mpc and its concentration parameter, $c\equiv R_{200}/r_{\rm s}$ (or $R_{\rm vir}/r_{\rm s}$) is $3.3$. The cluster is assumed to live on the cosmological background, so that the ratio of its effective and actual masses can be approximated as \citep[see, e.g.,][]{lzk2012}
\begin{eqnarray}\label{eq:etaz}
\eta(z)=\frac{M_{\rm halo,eff}}{M_{\rm halo,actual}} = \min\left\{1+\frac{\bar{f}_R(z)}{2\Phi_{\rm N}},\frac{4}{3}\right\}
\end{eqnarray}
where $\bar{f}_R(z)$ is the background value of $f_R$ at redshift $z$, and
\begin{eqnarray}
\Phi_{\rm N} = -\frac{GM_{\rm halo}}{R_{\rm halo}}\frac{\ln(1+c)}{\ln(1+c)-c/(1+c)} \approx -5\times10^{-5},
\end{eqnarray}
is the Newtonian potential at the edge of an NFW halo. We adopt the $f(R)$ model by \citet{hs2007} with $n=1$, for which
\begin{eqnarray}
\bar{f}_R(z) = \left[\frac{\Omega_m+4\Omega_\Lambda}{\Omega_m(1+z)^3+4\Omega_\Lambda}\right]^2\bar{f}_{R0}.
\end{eqnarray}
The values $\Omega_m=0.316$ and $\Omega_\Lambda=0.684$ are taken from the latest results of \citet{planck2015}. As a result, the physics of modified gravity is completely governed by $\bar{f}_{R0}$, which is the present-day value of $\bar{f}_R$. Once this is specified, we can obtain $\eta(z)$, and therefore infer $\left(\Omega_{\rm b}/\Omega_{\rm m}\right)_{\rm true}$ given $\left(\Omega_{\rm b}/\Omega_{\rm m}\right)_{\rm obs}$ and $z$ of a cluster.

In Fig.~\ref{fig:obom} we show the $\left(\Omega_{\rm b}/\Omega_{\rm m}\right)_{\rm true}$ result obtained from $f^{\rm obs}_{\rm gas}$ for 4 different cases: standard $\Lambda$CDM (upper left), and $f(R)$ gravity models with $|\bar{f}_{R0}|=3\times10^{-5}$ (upper right), $5\times10^{-5}$ (lower left) and $10^{-4}$ (lower right). For comparison, we have also, in each panel, plotted the mean value (solid line) and 1$\sigma$ confidence level (dotted) of $\Omega_{\rm b}/\Omega_{\rm m}$ from \citet{planck2015}\footnote{Note that the $f(R)$ models studied here have practically identical CMB power spectra as the $\Lambda$CDM model with the same $\Omega_{\rm m}$ and $\Omega_\Lambda$. As a result, using CMB data only, the constraints on cosmological parameters such as $\Omega_{\rm m}$ would be the same in all these models. Because of this, the CMB constraints are less model-dependent.}. The results from the 42 clusters, with 1$\sigma$ errors, are shown as symbols.

A quick naked-eye inspection shows that the $f_{\rm gas}$ method and the CMB observation give compatible $\Omega_{\rm b}/\Omega_{\rm m}$ if one assumes the $\Lambda$CDM paradigm (upper left). The $f(R)$ model with $|\bar{f}_{R0}|=10^{-4}$ (lower right), on the other hand, leads to a significantly higher value of $\Omega_{\rm b}/\Omega_{\rm m}$ than what CMB says, and is therefore inconsistent. The other two cases are more interesting: for $|\bar{f}_{R0}|=5\times10^{-5}$ (lower left), $\eta(z)$ increases to $4/3$ at $z\sim0.45$, while for $|\bar{f}_{R0}|=3\times10^{-5}$ (upper right) $\eta(z)$ only increases to $\sim1.28$ at $z=0$. In both cases, however, the inferred values of $\Omega_{\rm b}/\Omega_{\rm m}$ are still substantially larger than the Planck result, especially for the low-$z$ clusters. This shows that cluster gas fraction can be a potentially powerful test of gravity, using X-ray observations only. In such tests, lensing data can be a useful addition, but is not necessary.

\begin{figure}
 \includegraphics[width=88mm]{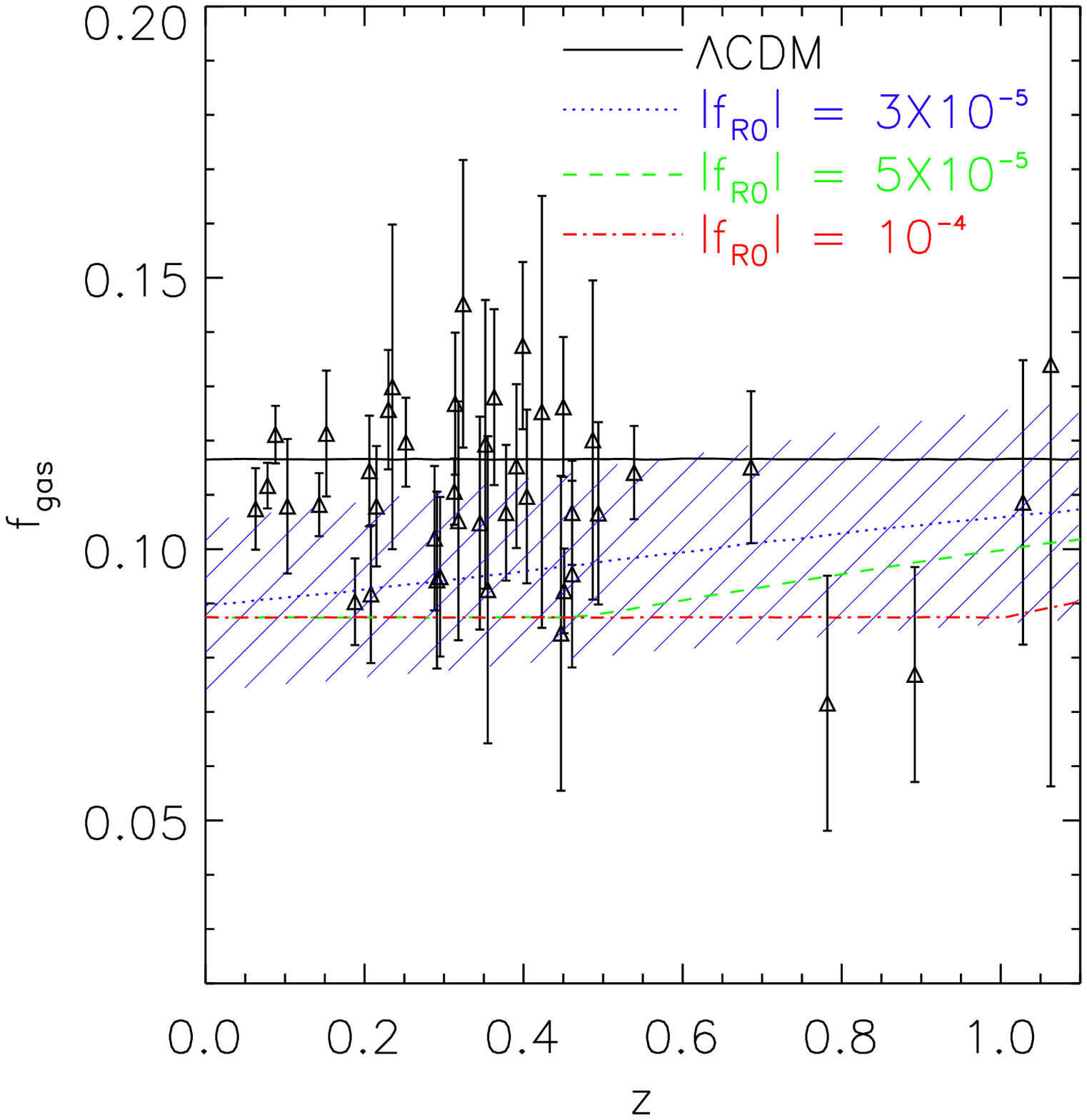}
\caption{(Colour online) The evolution of $f_{\rm gas}$ in 4 models with the same cosmic value of $\Omega_{\rm b}/\Omega_{\rm m}$ from \citet{planck2015}. The models are, respectively, $\Lambda$CDM (black solid line), and Hu-Sawicki $n=1$ $f(R)$ model with $|\bar{f}_{R0}|=3\times10^{-5}$ (blue dotted line), $5\times10^{-5}$ (green dashed line) and $10^{-4}$ (red dot-dashed line). Black triangles with error bars are the $f_{\rm gas}$ values of the 42 clusters used in \citet[][Tab 3]{allen2008}. The blue shaded region denotes the standard deviation around the mean $f_{\rm gas}$ for the model with $|\bar{f}_{R0}|=3\times10^{-5}$, for illustration, which shows that the theoretical uncertainty is roughly of the same order as current observational errors in $f_{\rm gas}$; therefore, the constraining power can be further improved if either of these uncertainties is reduced in the future.}
\label{fig:fgas_evol}
\end{figure}

The test can be done in an alternative way. For this, we assume the value of $\Omega_{\rm b}/\Omega_{\rm m}$ obtained from CMB observations, and check what value of $f^{\rm obs}_{\rm gas}$ an observer would have found  if living in a $f(R)$ universe. The idea is that, if this value differs too much from what {\it our} observers have told us \citep[e.g., in][]{allen2008}, then it would place a constraint on the extent to which the assumed $f(R)$ model can deviate from standard $\Lambda$CDM. As in the previous case, we draw random samples of size $10^5$ for the parameters $K, \gamma, b_0, \alpha_{\rm b}, s_0, \alpha_{\rm s}$ from which we find $10^5$ realisations of $K\gamma b(z)/(1+s(z))$. Then, by modelling modified gravity effects using Eq.~(\ref{eq:etaz}), we compute the mean $f_{\rm gas}$ for the 4 models shown in Fig.~\ref{fig:obom}, and these are shown as curves in Fig.~\ref{fig:fgas_evol} together with the observed values of $f_{\rm gas}$ from \citet{allen2008}. Again, we note that current data favour $\Lambda$CDM over all three variants of $f(R)$ gravity. As clusters are less screened at late times, we find that low-$z$ data is more useful in constraining the model than high-$z$ data.

\section{Discussion and Conclusions}

\label{sect:con}

In this paper, we proposed a new cosmological test of gravity, by inferring the cosmic baryon fraction from the apparent gas fractions of massive clusters, and comparing with the results from other, less model-dependent, measurements such as the CMB. In theories with a stronger gravity, the apparent gas fraction is smaller than that in $\Lambda$CDM for a fixed $\Omega_{\rm b}/\Omega_{\rm m}$. Reversely, if the observed value $f^{\rm obs}_{\rm gas}$ is fixed, we would find a higher value of $\Omega_{\rm b}/\Omega_{\rm m}$ than in GR, that can be inconsistent with the model-independent measurements. Taking the Hu-Sawicki $f(R)$ model as an example: our quick calculation shows that model parameters $|\bar{f}_{R0}|\sim3-5\times10^{-5}$ are in tension with the gas fraction data of the 42 clusters from \citet{allen2008}, though a more rigorous constraint will be left for future work.

$f_{\rm gas}$ has been a rather widly-used observable \citep[e.g.,][]{white1993}, and its power in constraining cosmology -- in particular dark energy models -- is convincingly demonstrated in various previous works \citep[e.g.,][]{sasaki1996,allen2004,allen2008}. The inclusion of baryons opens a new dimension for tests of gravity, since ultimately most cosmological observables can be tracked back to lights emitted by interactions involving baryons. In the mean time, the physics of the X-ray-emitting hot gas in massive clusters is relatively clean, making it easier both for the modelling and to use the observational data. As an example, the assumption that gas temperature depends on the gravitational potential and our main conclusions that (i) $f^\ast_{\rm gas}$ -- the true gas fraction -- is {\it unchanged} with modified gravity while (ii) $f^{\rm obs}_{\rm gas}$ is {\it changed} are supported by hydrodynamical simulations in $f(R)$ gravity \citep[e.g.,][]{he2015b}. Some uncertainties remain in relating $f_{\rm gas}$ to $\Omega_{\rm b}/\Omega_{\rm m}$, but these have been included in the error budget estimate above. Furthermore, within our current state of understanding, slightly changing its modelling (e.g., from \citet{allen2004} to \citet{allen2008}) does not change results drastically.

Here, we would like to emphasise the use of effective haloes \citep{he2015a} in our analysis. Though the idea has a similar origin as that of the dynamical mass of halo \citep[e.g.,][]{schmidt2010}, there are fundamental differences. Dynamical mass is a certain attribute of a given halo which is defined in the standard way, while effective halo is a completely new way to define and identify haloes. Given an effective halo, all gravitational effect can be calculated from GR, and in particular this means that the way in which $f_{\rm gas}$ is currently extracted from observational data -- and the resulting $f_{\rm gas}$ results -- can be directly used for our purpose. Thus, with a little extra effort from the people who generate a halo catalogue, the analyses of end users can be made much more straightforward, and this provides an efficient bridge between simulators, theorists and observers. 

One may naturally wonder about the generality of this method. As a cosmological test, it relies on galaxy clusters being totally or partially unscreened. Because we are talking about massive clusters which tend to be better screened, this test, {like most other cosmological ones}, will probably not be able to constrain $\bar{f}_{R0}$ to substantially smaller than the quoted values here. However, it does provide a fairly clean test -- with good observational data available -- that has the potential to place one of the strongest constraints from cosmology on $f(R)$ gravity. Also, one can always combine $f_{\rm gas}$ and other observables, such as lensing \citep{terukina2014}, cluster scaling relations \citep[e.g.][]{arnold2014}, and cluster gas pressure profiles \citep[][]{demartino2014}, to place joint, and likely stronger, constraints. In principle, the test would be more powerful if observational data for smaller galaxy clusters (e.g., those in the mass range $10^{13}\sim10^{14}h^{-1}M_\odot$) and galaxy groups are included, because these objects are less screened and so gravity deviates more from GR in general. This, however, requires a better understanding of the feedbacks in different models, which are not well studied so far.

The test can be applied not only to $f(R)$ gravity and the more general chameleon theory, but also to similar models such as dilatons \citep{dilaton} and symmetrons \citep{symmetron}. These models are all featured by a universal coupling of all matter species to a scalar field, that effectively enhances the gravity for all particles (at least in unscreened regimes). There are models in which only certain matter species, e.g.. dark matter, experiences the scalar coupling: therein, baryons can still feel a different gravity depending on how the dark matter particle mass evolves with time, in which case the proposed test does apply. In addition to these theories, we have mentioned above the DGP, Galileon and k-mouflage models. In the first two classes, the deviations from GR are strongly suppressed inside dark matter haloes \citep[][]{barreira2013,barreira2014b}, and so we do not expect the new test to work. For K-mouflage, as is for the so-called non-local gravity \citep{nonlocal1,nonlocal2,barreira2014a}, there can be a time evolution of Newton's constant inside clusters, making it possible to use this test. However, one needs to bear in mind that in many of these theories the background evolution history is also modified, and that can affect the $f_{\rm gas}$ test (whether it leads to degeneracies or stronger constraints can only be told by a case-by-case study in the future). 

As mentioned earlier, the aim of this paper is to illustrate the main idea of using $f_{\rm gas}$ as a test  of gravity theories, and therefore we have made a simplified estimate and have not quoted any numerical results on the confidence levels of the constrained $f_{R0}$. A more complete and rigorous analysis will require one to relax the simplification that all observed clusters share the same mass, radius and concentration, and use the real observational results of these for all clusters. If the cluster mass is obtained from its dynamical effects, we also need to account for the fact that different clusters may have experienced different degrees of screening, and so a more accurate modelling of the screening is needed to compute the cluster mass profile. These will be left for future work. We note that hydrodynamical simulations for modified gravity theories started to appear recently \citep[e.g.,][]{arnold2014,hammami2015,he2015b}, and such works will be useful for improving the constraining power of this test in the future.

\section*{Acknowledgments}

We thank Sownak Bose, Vince Eke, Claudio Llinares and Gongbo Zhao for discussions and comments. The work has used the DiRAC Data Centric system at Durham University, operated by the Institute for Computational Cosmology on behalf of the STFC DiRAC HPC Facility (www.dirac.ac.uk). This equipment was funded by BIS National E-infrastructure capital grant ST/K00042X/1, STFC capital grant ST/H008519/1, and STFC DiRAC Operations grant ST/K003267/1 and Durham University. DiRAC is part of the National E-Infrastructure.  BL acknowledges support by the UK STFC Consolidated Grant No.~ST/L00075X/1 and No.~RF040335. JHH is supported by the Italian Space Agency (ASI) via contract agreement I/023/12/0. LG acknowledges support from {NSFC} grants Nos. 11133003 and 11425312, the Strategic Priority Research Program {\it The Emergence of Cosmological Structure}  of the Chinese Academy of Sciences (No.~XDB09000000), {MPG} partner Group family, and an {STFC} and Newton Advanced Fellowship.

%\onecolumn
%\appendix
%
%\section{Beyond the First-order Approximations}

\label{lastpage}


\begin{thebibliography}{}
\bibitem[\protect\citeauthoryear{Allen~et~al.}{2004}]{allen2004} Allen S.~W., Schmidt R.~W., Ebeling H., Fabian A.~C., van Speybroeck L., 2004, MNRAS, 353, 457
\bibitem[\protect\citeauthoryear{Allen~et~al.}{2008}]{allen2008} Allen S.~W., Rapetti D.~A., Schmidt R.~W., Ebeling H., Morris R.~G., Fabian A.~C., 2008, MNRAS, 383, 879
\bibitem[\protect\citeauthoryear{Arnold, Puchwein \& Springel}{2014}]{arnold2014} Arnold C., Puchwein E., Springel V., 2014, MNRAS, 440, 833
\bibitem[\protect\citeauthoryear{Barreira~{et~al}.}{2013}]{barreira2013} Barreira A., Li B., Hellwing W.~A., Baugh C.~M., Pascoli S., 2013, JCAP, 10, 027
\bibitem[\protect\citeauthoryear{Barreira~{et~al}.}{2014a}]{barreira2014a} Barreira A., Li B., Hellwing W.~A., Baugh C.~M., Pascoli S., 2014, JCAP, 09, 031
\bibitem[\protect\citeauthoryear{Barreira~{et~al}.}{2014b}]{barreira2014b} Barreira A., Li B., Hellwing W.~A., Lombriser L., Baugh C.~M., Pascoli S., 2014b, JCAP, 04, 029
\bibitem[\protect\citeauthoryear{Brax \& Valageas.}{2014a}]{kmouflage1} Brax P., Valageas P., 2014a, PRD, 90, 023507
\bibitem[\protect\citeauthoryear{Brax \& Valageas.}{2014b}]{kmouflage2} Brax P., Valageas P., 2014b, PRD, 90, 023508
\bibitem[\protect\citeauthoryear{Brax~{et~al}.}{2011}]{dilaton} Brax P., van de Bruck C., Davis A.~C., Shaw D.~J., 2010, PRD, 82, 063519
\bibitem[\protect\citeauthoryear{Bose, Hellwing \& Li.}{2015}]{bose2015} Bose S., Hellwing W.~A., Li B., 2015, JCAP, 02, 034
\bibitem[\protect\citeauthoryear{Carroll~{et~al}.}{2005}]{cddatt2005} Carroll S.~M., de Felice A., Duvvuri V., Easson D.~A., Trodden M., Turner M.~S., 2005, PRD, 71, 063513
\bibitem[\protect\citeauthoryear{Cataneo~et~al.}{2014}]{cataneo2014} Cataneo M., Rapetti D., Schmidt F., Mantz A.~B., Allen S.~W.., Applegate D.~E., Kelly P.~L., von der Linden A., Morris R.~G., 2014, arXiv:1412.0133
\bibitem[\protect\citeauthoryear{Clifton~{et~al}.}{2012}]{clifton2012} Clifton T., Ferreira P.~G., Padilla A., Skordis C., 2012, Phys.~Rept., 513, 1
\bibitem[\protect\citeauthoryear{Copeland~{et~al}.}{2006}]{cst2006} Copeland E.~J., Sami M., Tsujikawa S., 2006, IJMPD, 15, 1753
\bibitem[\protect\citeauthoryear{De Martino~{et~al}.}{2014}]{demartino2014} De Martino I., De Laurentis M., Atrio-Barandela F., Capozzilello S., 2014, MNRAS, 442, 921
\bibitem[\protect\citeauthoryear{De Martino~{et~al}.}{2015}]{demartino2015} De Martino I., De Laurentis M., Capozzilello S., 2015, in {\it Special Issue Modified Gravity Cosmology: From Inflation to Dark Energy}; arXiv:1507.06123
\bibitem[\protect\citeauthoryear{Deffayet~{et~al}.}{2009}]{galileon2} Deffayet C., Esposito-Farese G., Vikman A., 2009, PRD, 79, 084003
\bibitem[\protect\citeauthoryear{Dirian et al.}{2014}]{nonlocal2} Dirian Y., Foffa S., Khosravi N., Kunz M., Maggiore M., 2014, JCAP, 06, 033
\bibitem[\protect\citeauthoryear{Dvali, Gabadadze \& Porrati}{2000}]{dgp} Dvali G., Gabadadze G., Porrati M., 2000, PLB, 485, 208
\bibitem[\protect\citeauthoryear{Eke, Navarro \& Frenk}{1998}]{eke1998} Eke V., Navarro J.~F., Frenk C.~S., 1998, ApJ, 503, 569
%\bibitem[\protect\citeauthoryear{Faulkner~{et~al.}}{2007}]{ftbm2007} Faulkner T., Tegmark M., Bunn E.~F., Mao Y., 2007, PRD, 76, 063505
\bibitem[\protect\citeauthoryear{Hammami et~al.}{2015}]{hammami2015} Hammami A., Llinares C., Mota D.~F., Winther, H. A., 2015, MNRAS, 449, 3635
\bibitem[\protect\citeauthoryear{He et~al.}{2015a}]{he2015a} He J.-h., Hawken A.~J., Li B., Guzzo L., 2015a, PRL, in press
\bibitem[\protect\citeauthoryear{He \& Li}{2015b}]{he2015b} He J.-h., Li B., 2015b, in preparation
\bibitem[\protect\citeauthoryear{Hinterbichler \& Khoury}{2010}]{symmetron} Hinterbichler K., Khoury J., 2010, PRL, 104, 231301
\bibitem[\protect\citeauthoryear{Hu \& Sawicki}{2007}]{hs2007} Hu W., Sawicki I., 2007, PRD, 76, 064004
\bibitem[\protect\citeauthoryear{Joyce~{et~al}.}{2015}]{joyce2015} Joyce A., Jain B., Khoury J., Trodden M., 2015, Phys.~Rept., 568, 1
\bibitem[\protect\citeauthoryear{Khoury \& Weltman}{2004}]{chameleon} Khoury J., Weltman A., 2004, PRD, 69, 044026
\bibitem[\protect\citeauthoryear{Komatsu \& Seljak}{2001}]{komatsu2001} Komatsu E., Seljak U., 2001, MNRAS, 327, 1353
%\bibitem[\protect\citeauthoryear{Lacey \& Cole}{1993}]{lc1993} Lacey C., Cole S., 1993, MNRAS, 262, 627
%\bibitem[\protect\citeauthoryear{Lahav~{et~al.}}{1991}]{llpr1991} Lahav O., Lilje P.~B., Primack J.~R., Rees M.~J., 1991, MNRAS, 251, 128
%\bibitem[\protect\citeauthoryear{Li}{2011}]{li2011} Li B., 2011, MNRAS, 411, 2615
%\bibitem[\protect\citeauthoryear{Li \& Barrow}{2007}]{lb2007} Li B., Barrow J.~D., 2007, PRD, 75, 084010
%\bibitem[\protect\citeauthoryear{Li \& Barrow}{2011}]{lb2011} Li B., Barrow J.~D., 2011, PRD, 83, 024007
%\bibitem[\protect\citeauthoryear{Li \& Efstathiou}{2012}]{le2012} Li B., Efstathiou G., 2012, MNRAS, 412, 1431.
%\bibitem[\protect\citeauthoryear{Li \& Lam}{2012}]{ll2012a} Li B., Lam T. Y., 2012, MNRAS, in press; arXiv:1205.0058 [astro-ph.CO].
\bibitem[\protect\citeauthoryear{Li, Zhao \& Koyama}{2012}]{lzk2012} Li B., Zhao G.-B., Koyama K., 2012, MNRAS, 421, 3481
%\bibitem[\protect\citeauthoryear{Li \& Zhao}{2009}]{lz2009} Li B., Zhao H., 2009, PRD, 80, 044027
%\bibitem[\protect\citeauthoryear{Li \& Zhao}{2010}]{lz2010} Li B., Zhao H., 2010, PRD, 81, 104047
%\bibitem[\protect\citeauthoryear{Li~{et~al.}}{2011}]{lztk2011} Li B., Zhao G., Teyssier R., Koyama K., 2011, arXiv:1110.1379 [astro-ph.CO]
\bibitem[\protect\citeauthoryear{Lombriser}{2014}]{lombriser2014} Lombriser L., 2014, Annalen der Physik, 526, 259
\bibitem[\protect\citeauthoryear{Maggiore \& Mancarella}{2014}]{nonlocal1} Maggiore M., Mancarella M., 2014, PRD, 90, 023005
\bibitem[\protect\citeauthoryear{Mo, van den Bosch \& White}{2011}]{mvw-book} Mo H.~J., van den Bosch F., White S.~D.~M., 2011, {\it Galaxy Formation and Evolution}, Cambridge University Press
\bibitem[\protect\citeauthoryear{Navarro, Frenk \& White}{1997}]{nfw} Navarro J.~F., Frenk C.~S., White S.~D.~M., 1997, ApJ, 490, 493
\bibitem[\protect\citeauthoryear{Nicolis~{et~al}.}{2009}]{galileon1} Nicolis A., Rattazzi R., Trincherini E., 2009, PRD, 79, 064036
%\bibitem[\protect\citeauthoryear{Oyaizu}{2008}]{oyaizu2008} Oyaizu H., 2008, PRD, 78, 123523
%\bibitem[\protect\citeauthoryear{Oyaizu~{et~al.}}{2008}]{olh2008} Oyaizu H., Lima M., Hu W., 2008, PRD, 78, 123524
\bibitem[\protect\citeauthoryear{Perlmutter~{et~al.}}{1999}]{perlmutter1999} Perlmutter S.~{\it et.~al.}, 1999, ApJ, 517, 565
\bibitem[\protect\citeauthoryear{Planck Collaboration, Ade~et~al.}{2015}]{planck2015} Planck Collaboration. Ade P.~A.~R., Aghanim N., Arnaud M., et al., 2015, arXiv:1502.01589
\bibitem[\protect\citeauthoryear{Riess~{et~al.}}{1998}]{riess1998} Riess A.~G.~{\it et.~al.}, 1998, Astron.~J., 116, 1009
\bibitem[\protect\citeauthoryear{Sasaki}{1996}]{sasaki1996} Sasaki, S., 1996, PASJ, 48, L119
\bibitem[\protect\citeauthoryear{Schmidt}{2010}]{schmidt2010} Schmidt F., 2010, PRD, 81, 103002
\bibitem[\protect\citeauthoryear{Schmidt, Vikhlinin \& Hu}{2009}]{schmidt2009} Schmidt F., Vikhlinin A., Hu W., 2009, PRD, 80, 083505
%\bibitem[\protect\citeauthoryear{Schaeffer \& Silk}{1988}]{ss1988} Schaeffer R., Silk J., 1988, ApJ, 292, 319
%\bibitem[\protect\citeauthoryear{Sheth}{1998}]{sheth1998} Sheth R.~K., 1998, MNRAS, 300, 1057
%\bibitem[\protect\citeauthoryear{Sheth \& Tormen}{2002}]{st2002} Sheth R.~K., Tormen G., 2002, MNRAS, 329, 61
%\bibitem[\protect\citeauthoryear{Sheth \& van de Weygaert}{2004}]{sv2004} Sheth R.~K., van de Weygaert R., 2004, MNRAS, 350, 517
\bibitem[\protect\citeauthoryear{Terukina {et~al.}}{2014}]{terukina2014} Terukina A., Lombriser L., Yamamoto K., Bacon D., Koyama K., Nichol R.~C. 2014, JCAP, 04, 013
%\bibitem[\protect\citeauthoryear{Valageas}{2009}]{valageas2009} Valageas P., 2009, A~\&~A, 508, 93
\bibitem[\protect\citeauthoryear{White {et~al.}}{1993}]{white1993} White S.~D.~M., Navarro J.~F., Evrard A.~E., Frenk C.~S., 1993, Nature, 366, 429
\bibitem[\protect\citeauthoryear{Wilcox {et~al.}}{2015}]{wilcox2015} Wilcox H., Bacon D., Nichol R.~C., et al., 2015, arXiv:1504.03937
%\bibitem[\protect\citeauthoryear{Will}{2006}]{cw2006} Will C.~M., {\it "The Confrontation between General Relativity and Experiment"}, Living Rev. Relativity 9,  (2006),  3. URL: http://www.livingreviews.org/lrr-2006-3
\bibitem[\protect\citeauthoryear{Zhao, Li \& Koyama}{2011}]{zlk2011} Zhao G.-B., Li B., Koyama K., 2011, PRL, 107, 071303
\end{thebibliography}
\end{document}